\documentstyle[12pt,aaspp4]{article}
\eqsecnum
\begin{document}
\newcommand{\half}{\case{1}{2}}
\newcommand{\Aa}{{\cal A}}
\newcommand{\Ab}{\partial{\cal A}/\partial\eta}
\newcommand{\Ac}{\partial^2{\cal A}/\partial\eta^2}
\newcommand{\Ba}{{\cal B}}
\newcommand{\kmpp}{\kappa_m^{\prime\prime}}
\title{Relativistic Diskoseismology. I. \\ Analytical Results for `Gravity
Modes'}
\author{Christopher A. Perez\altaffilmark{1},
Alexander S. Silbergleit\altaffilmark{2},
Robert V. Wagoner\altaffilmark{3}, \\ and Dana E. Lehr\altaffilmark{4}}
\affil{Department of Physics and Center for Space Science and Astrophysics
\\ Stanford University, Stanford, CA 94305--4060}

\altaffiltext{1}{Present address: Morgan Stanley \& Co., 1585 Broadway, 
New York, NY 10036; Internet: perezc@ms.com}
\altaffiltext{2}{Also Gravity Probe--B, Stanford University; Internet:
gleit@relgyro.stanford.edu}
\altaffiltext{3}{Internet: wagoner@leland.stanford.edu}
\altaffiltext{4}{Internet: danalehr@leland.stanford.edu
\\[0.5cm] SU--ITP--96/1. Submitted to The Astrophysical Journal 
on June 14, 1996.}

\begin{abstract}

We generalize previous calculations to a fully relativistic treatment of
adiabatic oscillations which are trapped in the inner regions of accretion
disks by non-Newtonian gravitational effects of a black hole. We employ the
Kerr geometry within the scalar potential formalism of Ipser and Lindblom,
neglecting the gravitational field of the disk. This approach treats
perturbations of arbitrary stationary, axisymmetric, perfect fluid models.
It is applied here to thin accretion disks. Approximate analytic eigenfunctions and eigenfrequencies are obtained for 
the most robust and observable class of modes, which corresponds roughly to the gravity (internal) oscillations of stars. The dependence of the oscillation frequencies on the mass and angular momentum of the black hole is exhibited. These trapped modes do not exist in Newtonian gravity, and thus provide a signature and probe of the strong-field structure of black holes. Our predictions are relevant to observations which could detect modulation of the X-ray luminosity from stellar mass black holes in our galaxy and the UV and optical luminosity from supermassive black holes in active galactic nuclei.

\end{abstract}


\newpage
\section{INTRODUCTION}

Our group has been investigating consequences of the realization
[originally by Kato \& Fukue \markcite{KF} (1980)] that general relativity (or similar relativistic theories of gravitation) can trap normal modes of oscillation in the regions of accretion disks near black holes (or weakly magnetized neutron stars). Since no such trapped modes exist in Newtonian gravity, they provide a signature for the existence of strong gravitational fields. The frequencies of these modes serve as probes of the structure of the field as well as the physical conditions within the accretion disk.

Using a modified Newtonian potential, adiabatic eigenfunctions and
eigenvalues of the lowest acoustic ($p$) modes \markcite{NW1} (Nowak \& Wagoner 1991, hereafter NW1) and internal gravity ($g$) modes \markcite{NW2} (Nowak \& Wagoner 1992, hereafter NW2) have been calculated. Their rates of growth or damping due to anisotropic (turbulent) viscosity and gravitational radiation reaction were also determined (NW2). Finally, the amplitude and coherence time of the oscillations, and the resulting fractional modulation of the spectral luminosity of the disk, have been estimated \markcite{NW3} (Nowak \& Wagoner 1993, hereafter NW3). Although this modulation appears to be at most a few percent, we feel that the potential payoff certainly justifies dedicated monitoring of a variety of supermassive black hole (AGN) candidates as well as the appropriate binary X-ray sources in our galaxy.

In this paper we begin to reanalyze such modes of oscillation in terms of
perturbations of the general relativistic equations of motion of perfect
fluids within the Kerr metric.
Ipser \& Lindblom \markcite{IL} (1992, hereafter IL) have developed an elegant formalism to analyze small adiabatic pulsations of relativistic perfect fluids which in equilibrium are in pure rotation in stationary and axisymmetric spacetimes. We apply this formalism to a thin accretion disk around a Kerr black hole. In this case, the Cowling approximation of negligible self-gravitational effects of the disk should be valid. All perturbed quantities can then be obtained from a scalar potential governed by a single linear two-dimensional second-order partial differential equation. With the appropriate boundary conditions, this equation yields trapped modes with discrete spectra which in combination can determine the black hole mass $M$ and dimensionless angular momentum parameter $a = cJ/GM^2$.

We then focus on the analogs of the gravity modes that exist within stars. They should be the most observable trapped modes since they appear to occupy the largest area of the disk, including its maximum temperature. Approximate expressions for their eigenfunctions and eigenfrequencies are obtained, and explicit results are presented for the case of small buoyancy frequency. 

\pagebreak[3]
\section{PERTURBATIONS OF BLACK HOLE ACCRETION DISKS}

\subsection{General Relations Within Thin Disks}

Unless otherwise indicated, we will adopt the conventions and notation of
IL (such as gravitational units: $G=c=1$). In addition, all lengths and
times are made dimensionless by expressing them in units of the black hole
mass $M$.
The matter within the accretion disk is assumed to be represented by a
perfect-fluid stress-energy tensor:
\begin{equation}
T^{ab} = \rho u^a u^b + pq^{ab} \; ,
\end{equation}
where $\rho$ is the mass-energy density, $p$ is the pressure, $u^a$ is the
four-velocity of the fluid (with $u^a u_a = -1$), and the operator $q^{ab}
= g^{ab} + u^a u^b$ projects components perpendicular to $u^a$. In the
neighborhood of the equatorial plane $z=0$, we use the Novikov and Thorne
\markcite{NT} (1973) expression for the Kerr metric in Boyer-Lindquist coordinates :
\begin{eqnarray}
ds^2 & = & -\frac{r^2\Delta^*}{A^*}dt^2 + \frac{A^*}{r^2}(d\phi -
\tilde{\omega}dt)^2
     + \frac{r^2}{\Delta^*}dr^2 + dz^2 \\
     \left(\frac{d}{ds}\right)^2 & = &
     - \frac{A^*}{r^2\Delta^*} \left(\frac{\partial}{\partial t}
     + \tilde{\omega}\frac{\partial}{\partial\phi} \right)^2
     + \frac{r^2}{A^*} \left( \frac{\partial}{\partial \phi} \right)^2
     + \frac{\Delta^*}{r^2} \left( \frac{\partial}{\partial r} \right)^2
     + \left( \frac{\partial}{\partial z} \right)^2 \; ,
\end{eqnarray}
with
\begin{equation}
\Delta^* \equiv r^2 - 2r + a^2 \; , \;\; A^* \equiv r^4 + r^2a^2 + 2ra^2 \;
, \;\; \tilde{\omega} \equiv 2ar/A^* \; ,
\end{equation}
where corrections of order $(z/r)^2$ are neglected (but not later, in the
vertical equations of motion). Note that $\| g_{ab} \| = -r^2$ and $g_{tt}
= -\Delta^*g^{\phi\phi} = -1+2/r$.

We further assume that $|u^z| \ll |u^r| \ll |ru^{\phi}|$ within the
stationary and axisymmetric unperturbed disk, in which the properties of
the fluid are also symmetric about the equatorial plane. The fluid
particles are thus travelling in approximately circular orbits (assumed in
corotation with the black hole) with four-velocity
\begin{equation}
u^a = \beta (t^a + \Omega\varphi^a) \; .
\end{equation}
With $x^a=\{t,r,\phi,z\}$, the Killing vector components are simply $t^a =
(1,0,0,0)$ and $\varphi^a = (0,0,1,0)$. Within the unperturbed disk the
Euler equation, in the form $q_{ac}\nabla_bT^{bc} = 0$, becomes
\begin{equation}
u^b\nabla_b u_a = -(\nabla_a\beta)/\beta + \beta\varphi^bu_b\nabla_a\Omega
= -(\nabla_a p)/\rho \; ,
\end{equation}
where $\nabla_a$ is the covariant derivative.
We have set $\rho + p \approx \rho$ since the speed of sound $c_s \sim
h\Omega \ll r\Omega < c = 1$, where $h(r)$ is the effective half-thickness
of the disk.

The functions $\beta$ (equivalent to the function $\gamma$ employed by IL)
and $\Omega$ (the angular velocity measured by an observer far from the
disk) are well approximated by the free-particle expressions
\begin{equation}
\beta = \frac{r^{3/2} + a}{r^{3/4}(r^{3/2}-3r^{1/2}+2a)^{1/2}} \;\; , \;\;
\Omega = \frac {1}{r^{3/2}+a} \; ,
\end{equation}
since the disk is assumed thin.
Away from the inner edge of thin, and even slim \markcite{A} (Abramowicz et al. 1988) disks, the effects of the actual radial velocity and radial pressure gradient on our eigenmodes will be small (NW3).

Next, we consider the linearized equations for the Eulerian (at a fixed
point in spacetime) perturbations $\delta\rho$, $\delta p (\ll
\delta\rho)$, and $\delta u^a$. We obtain these by perturbing the two
projections of the energy-momentum conservation equation, the law of energy
conservation
\begin{eqnarray}
u_b \nabla_a T^{ab} = 0 & \Rightarrow & 0 = u^a \nabla_a \delta \rho
     + \delta \hat{u}^a \nabla_a \rho + \rho \nabla_a \delta \hat{u}^a \; ,
     \label{eq:Energy} \\*
\noalign{\noindent and the Euler equation of motion} \nonumber \\*
q^a_c \nabla_b T^{bc} = 0 & \Rightarrow & 0 = \rho (q^a_b u^c \nabla_c
     \delta \hat{u}^b + \delta \hat{u}^b \nabla_b u^a)
     + \delta \rho u^b \nabla_b u^a + q^{ab} \nabla_b \delta p \; ,
     \label{eq:Euler}
\end{eqnarray}
where $\delta \hat{u}^a = q^a_b \delta u^b$ and $u_a \delta u^a = 0$.

We impose the adiabatic condition that the perturbations preserve the
entropy per baryon ($s$). This condition implies the following relationship
between Lagrangian variations (associated with a given element of fluid) in
the pressure and the density:
\begin{equation}
\Delta p = (\Gamma p/\rho) \Delta \rho \; .            \label{eq:thermo}
\end{equation}
Although IL generalized $\Gamma$ to be an arbitrary stationary and
axisymmetric function in the equilibrium fluid, for these adiabatic
perturbations $\Gamma = (\rho/p)(\partial p /\partial \rho)_s =
(\rho/p)c_s^2 $. Under most conditions (gas or radiation pressure
dominance), the equation of state will be of the form $p =
K(s)\rho^\Gamma$, so we shall take the adiabatic index $\Gamma$ (denoted
$\gamma$ by NW1--3) to be at most a slowly varying function of $r$.

One can relate the Lagrangian and Eulerian variations via the Lagrangian
displacement vector $\xi^a$, giving $\Delta p = \delta p + \xi^a \nabla_a
p$ and $\Delta\rho = \delta\rho + \xi^a \nabla_a \rho$. The negative Lie
derivative of the unperturbed velocity gives its Eulerian perturbation
\markcite{FS} (Friedman \& Schutz 1978): $\delta\hat{u}^a = q^a_b (u^c \nabla_c \xi^b - \xi^c \nabla_c u^b)$.
The previous equation implies a gauge freedom in $\xi^a$, corresponding to
the addition to $\xi^a$ of any function parallel to $u^a$ \markcite{SS} (Schutz \& Sorkin 1977). We choose $\xi^a u_a = 0$ to fix the gauge.

From now on, we shall work with the azimuthal and temporal Fourier
components of all quantities such as the displacement vector $\xi^a
\rightarrow \xi^a(r,z)\exp[i(m\phi + \sigma t)]$. We also employ the
corotating frequency $\omega(r) = \sigma + m\Omega(r)$. (Note that IL
exchange the definitions of $\omega$ and $\sigma$; we employ those of
NW1--3.)  This Fourier decomposition provides us with an algebraic
relationship between $\xi^a$ and $\delta\hat{u}^a$:
\begin{equation}
\xi^a = -i(\omega\beta)^{-1}[q^a_b - (i/\omega)q^a_c\varphi^c\nabla_b
\Omega]\delta\hat{u}^b \; .                                \label{eq:xi}
\end{equation}

With $p \ll \rho$ as usual, the scalar potential employed by IL becomes
$\delta U \equiv \delta p/\rho$, the specific adiabatic Eulerian
perturbation of the enthalpy. However, we will also find it convenient to
use their rescaled potential
\begin{equation}
\delta V \equiv \delta U/(\beta\omega) = \delta p/(\rho\beta\omega) \; .
\label{eq:deltaV} \end{equation}
Following IL, one can then obtain the relation
\begin{equation}
\delta\rho = \rho^2\left(\frac{\delta U}{p\Gamma}
+ \frac{iA_a\delta\hat{u}^a}{\omega\beta}\right)        \label{eq:delrho}
\end{equation}
in terms of the function
\begin{equation}
A_a = \frac{\nabla_a \rho}{\rho^2} - \frac{\nabla_a p}{\rho p\Gamma}
    = \rho^{-1}\nabla_a{\cal A} \; ,                         \label{eq:A}
\end{equation}
where the final equality is valid if $\Gamma$ is constant, with ${\cal A}
\equiv \ln(\rho/p^{1/\Gamma}) + $ constant. Note [from equation
(\ref{eq:thermo})] that $A_a$ vanishes for any barotropic equation of state
$p=p(\rho)$.

Since the accretion disks that we shall consider are thin, vertical
gradients of equilibrium quantities (such as pressure) are typically much
greater than their radial gradients. Therefore the Brunt--
V\"{a}is\"{a}l\"{a} or buoyancy frequency, $N_z$, is given by
\begin{equation}
A^a\nabla_a p \cong A^z\nabla_z p \equiv \beta^2 N_z^2 \; .  \label{eq:B}
\end{equation}

Radial perturbations of the circular orbits of free particles
(corresponding to a zero-pressure, infinitesimally thin disk) oscillate at
the epicyclic frequency $\kappa$ given by \markcite{OKF} (Okazaki, Kato, \& Fukue 1987)
\begin{equation}
2\omega^a\Omega_a \cong 2\omega^z\Omega_z = \beta^2\kappa^2 =
\beta^2\Omega^2
     \left(1-\frac{6}{r}+\frac{8a}{r^{3/2}}-\frac{3a^2}{r^2}\right) \; ,
\end{equation}
where $\omega^a$ (not to be confused with the corotating frequency
$\omega$) and $\Omega^a$ are the vorticity and angular velocity vectors
(IL).
Similarly, vertical perturbations of free particle circular orbits produce oscillations at the vertical frequency $\Omega_\perp$ given by \markcite{K0} (Kato 1990)
\begin{equation}
\Omega_\perp^2 =
\Omega^2\left(1-\frac{4a}{r^{3/2}}+\frac{3a^2}{r^2}\right) \; .
\end{equation}
The frequency $\Omega_\perp$ also appears in the equation of vertical hydrostatic equilibrium \markcite{K3} (Kato 1993):
\begin{equation}
\partial p/\partial z = -\rho\beta^2\Omega_\perp^2 z \; ,
\label{eq:hydrostatic}
\end{equation}
and therefore in our definition of disk half-thickness h(r):
\begin{equation}
p(r,0)/\rho(r,0) \equiv (h\beta\Omega_\perp)^2 \; .
\label{eq:defh}
\end{equation}
We note that equation (\ref{eq:hydrostatic}) is incorrect in Novikov \&
Thorne \markcite{NT} (1973), as also pointed out by Riffert \& Herold \markcite{RH} (1995). We can now
obtain the useful expression
\begin{equation}
N_z^2 = -\Omega_\perp^2(z\partial{\cal A}/\partial z) \; ,\label{eq:buoy}
\end{equation}
from equations (\ref{eq:A}) and (\ref{eq:B}).

\begin{figure}
\plotone{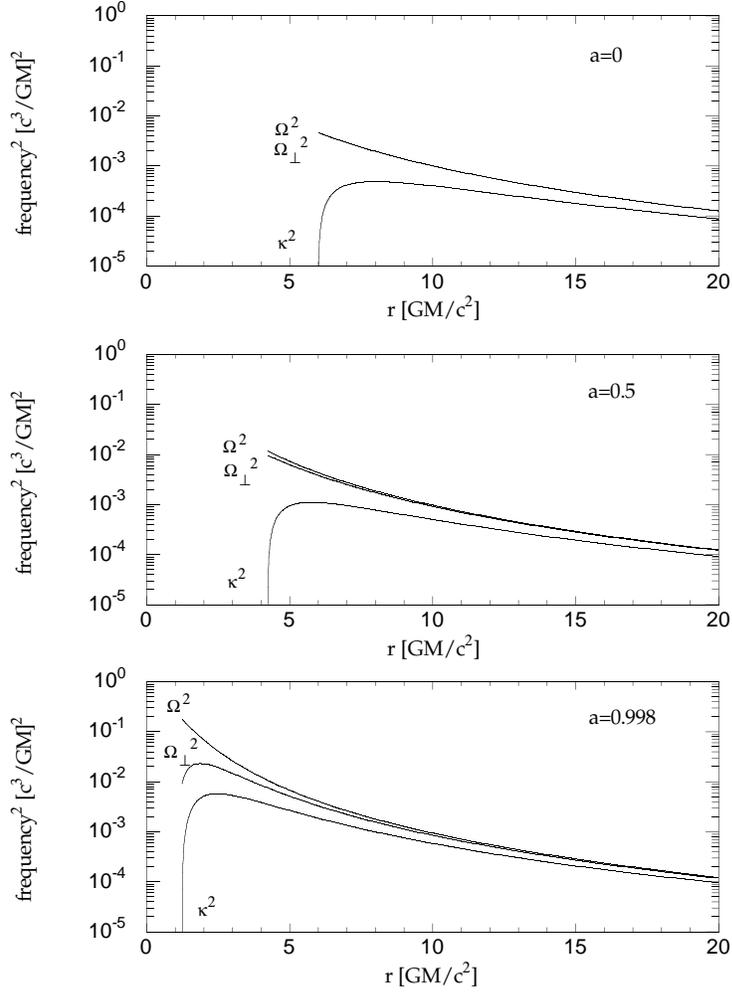}
\caption{The radial dependence of the square of the fundamental free-particle frequencies that govern the modes of the disk: Keplerian ($\Omega$), and radial ($\kappa$) and vertical ($\Omega_\perp$) epicyclic. Three values of the black hole angular momentum parameter $a=cJ/GM^2$ are chosen.}
\label{orbital}
\end{figure}

These frequencies ($\kappa$, $\Omega_\perp$, and $N_z$) play the key role
in determining the modes of oscillation of the disk. In
Figure~\ref{orbital}, we show the radial dependence of the free-particle
orbital frequencies $\Omega$, $\kappa$, and $\Omega_\perp$, for three values of the angular momentum parameter $a$. Thorne \markcite{T} (1974) has shown that a black hole can be spun up by accretion to a value of $a=0.998$, only slightly less than its absolute limit of unity.

The equations which govern the perturbations of the disk simplify
considerably if the perturbations of the gravitational field are negligible
(Cowling approximation), as is true for thin accretion disks. Using the
perturbed equation of motion (\ref{eq:Euler}), IL are then able to
algebraically relate the Eulerian perturbations $\delta u^a$, $\delta
\rho$, and $\delta p$ to $\delta U$ and its covariant derivative. Using the
perturbed equation of energy conservation (\ref{eq:Energy}), IL then obtain
a two-dimensional partial differential `master' equation [their equation
(39)] which governs the rescaled potential $\delta V$. For our problem it
assumes the form
\begin{equation}
D_\mu\left[(\Delta^*)^{1/2}\rho H^{\mu\nu}D_\nu\delta V\right] +
(\Delta^*)^{1/2}\omega\beta\Phi\delta V = 0 \; ,
\label{eq:master1}
\end{equation}
where $D_\mu$ ($\mu = r,z$) is the covariant derivative associated with the
metric $g^{\mu\nu}$ of the plane orthogonal to the Killing vectors.
We find (from the expressions in IL) that
\begin{equation}
H^{rr} = g^{rr}\omega^2(\omega^2-\kappa^2)^{-1} \quad , \quad
H^{zz} = \omega^2(\omega^2-N_z^2)^{-1} \; ,
\end{equation}
and $H^{rz}$ is negligible. In addition, we obtain
\begin{equation}
\Phi = \rho\omega\beta c_s^{-2} - \rho^3(\omega\beta)^{-1}H^{zz}A_z^2
+ \nabla_z[\rho^2(\omega\beta)^{-1}H^{zz}A_z] \; .   \label{eq:Phi}
\end{equation}
We have restricted consideration to low azimuthal modes: $|m| \ll r/h$. The
reduction of the function $\Phi$ to the above form also depends upon the
assumption that the corotating eigenfrequency $\omega$ lies in the range
$h/r \ll |\omega|/\Omega \ll r/h$.

\pagebreak[3]
\subsection{Radial-WKB Approximation and Separation of Variables}

IL applied their potential formalism to two problems: general adiabatic
perturbations of nonrotating stars and barotropic perturbations of rigidly
rotating stars. As in NW2,3, we shall apply it to thin disk  perturbations
whose radial wavelength $\lambda(r,z) \equiv 2\pi|\delta V(\partial\delta
V/\partial r)^{-1}|$ is much less than $r$, the scale of radial variation of equilibrium properties of the disk. This turns out to be a good
approximation, and it was also found in NW1--3 that $\lambda \sim h(r)$ for
most of the lowest modes studied. Therefore, we work to lowest nontrivial
order in the WKB parameter $\lambda/r$. The master equation
(\ref{eq:master1}) then becomes
\begin{equation}
g^{rr}\frac{\partial}{\partial r}\left[\frac{1}{(\omega^2-\kappa^2)}
\frac{\partial\delta V}{\partial r}\right] +
\frac{1}{\rho}\frac{\partial}{\partial z}\left[\frac{\rho}{(\omega^2-
N_z^2)}\frac{\partial\delta V}{\partial z}\right] + \left[
\frac{\beta^2}{c_s^2} + \frac{1}{p^{1/\Gamma}}\frac{\partial}{\partial z}
\left\{\frac{p^{1/\Gamma}}{(\omega^2-N_z^2)}\frac{\partial {\cal
A}}{\partial z}\right\}\right]\delta V = 0 \; .
\label{eq:master2}
\end{equation}
We have also allowed for rapid variations near the radial corotation
resonance at $\omega^2 = \kappa^2$. In the quasi-Newtonian limit, NW2
neglected the terms involving $\partial(\omega^2-N_z^2)/ \partial z$ in
their corresponding equation (2.4).

In order to express Lagrangian perturbations such as the boundary condition
$\Delta p = \omega\beta\rho \delta V + \xi^a\nabla_a p = 0$, we need the
expressions
\begin{equation}
\xi^r \cong \frac{\omega g^{rr}}{\beta(\omega^2-\kappa^2)}
\frac{\partial}{\partial r}\delta V \quad , \quad \xi^z \cong
\frac{\omega}{\beta(\omega^2-N_z^2)}\left[\frac{\partial}{\partial z}
\delta V + \rho A_z \delta V\right]             \label{eq:xirz}
\end{equation}
for the relevant components of the displacement vector. We also find that
\begin{equation}
\xi^\phi \cong -i\frac{(-u^tu_t)}{\omega}\left(\frac{\partial\Omega}
{\partial r} + \frac{r\omega^z}{\beta^2\Delta^*}\right)\xi^r \; ,
\label{eq:xiphi}
\end{equation}
where $u^tu_t = -\beta^2[1-(2/r)(1-a\Omega)]$.

As in previous work (NW2,3), equation (\ref{eq:master2}) is separable in
$r$ and $\eta \equiv z/h(r)$ to lowest WKB order, if we assume that
$c_s^2(r,z) \equiv c_s^2(r,0)g(r,\eta)$ is a slowly-varying function of
$r$. We note, from our definition (\ref{eq:defh}) of $h(r)$, that
$c_s^2(r,0) = \Gamma(h\beta \Omega_\perp)^2$. In order to bring the radial
separated equation to a familiar form [cf.~short wavelength perturbations
of nonrotating stellar atmospheres \markcite{H} (Hines 1960)], we take the slowly varying separation function to be
\begin{displaymath}
\frac{1}{\Gamma(h\omega)^2}\left[\Psi(r) -
\frac{\omega^2}{\Omega_\perp^2}\right] \; .
\end{displaymath}
The redefined separation function $\Psi(r)$ was denoted $\Gamma\Upsilon(r)$
by NW2,3. So now we set $\delta V = V_r(r) V_\eta(r,\eta)$, with $V_\eta$ a
slowly-varying function of $r$, and obtain the separated equations
\begin{equation}
\frac{d^2 V_r}{dr^2} - \frac{1}{(\omega^2-\kappa^2)} \frac{d}{dr}
(\omega^2-\kappa^2)\frac{dV_r}{dr} +\alpha^2(\omega^2-\kappa^2)(1 -
\Psi/\omega_*^2)V_r = 0 \; , \label{eq:sep1}
\end{equation}
{\samepage \begin{displaymath}
\frac{\partial^2V_\eta}{\partial\eta^2} - \left[\frac{\eta}{\Gamma g} -
\frac{\partial{\cal A}}{\partial\eta} + \Xi\right]\frac{\partial
V_\eta}{\partial\eta}
\end{displaymath}
\begin{equation}
+ \left[\frac{\omega_*^2}{\Gamma g} +
\frac{(\Psi - \omega_*^2)}{\Gamma}\left(1 +
\frac{\eta}{\omega_*^2}\frac{\partial{\cal A}}{\partial\eta}\right) +
\frac{\partial^2{\cal A}}{\partial\eta^2} - \Xi\frac{\partial{\cal
A}}{\partial\eta}\right]V_\eta = 0 \; ,       \label{eq:sep2}
\end{equation} }
where $\omega_* \equiv \omega/\Omega_\perp$ and
\begin{equation}
\alpha^2 \equiv \frac{\beta^2 g_{rr}}{c_s^2(r,0)}
= \frac{g_{rr}}{\Gamma h^2 \Omega_\perp^2} \; , \quad
\Xi \equiv {\cal R}\left(\eta\frac{\partial^2{\cal A}}{\partial\eta^2} +
\frac{\partial{\cal A}}{\partial\eta}\right)  \; , \quad
{\cal R} \equiv \frac{\Omega_\perp^2}{(\omega^2-N_z^2)} \; .
\label{eq:aux1}
\end{equation}
Equations (\ref{eq:sep1}) and (\ref{eq:sep2}) agree with equations (2.3a)
and (2.3b) of NW3 in the quasi-Newtonian limit, except for the term
proportional to $d(\omega^2-\kappa^2)/dr$ missing in NW3.

\pagebreak[3]
\subsection{Boundary Conditions}

We shall take the top and bottom surfaces of the disk [$\eta =
\pm\eta_0(r)$] to be defined by the vanishing of the equilibrium pressure
$p$. The boundary condition on the perturbations there is taken to be the
usual one: $0 = \Delta p \cong \delta p + \xi^z \nabla_z p$. Using
equations (\ref{eq:deltaV}), (\ref{eq:hydrostatic}), (\ref{eq:buoy}),
(\ref{eq:xirz}), and (\ref{eq:aux1}), this becomes
\begin{equation}
\Delta p = \left(\frac{\beta\omega\rho}{\omega_*^2 +
\eta\Ab}\right)\Ba(\delta V) \rightarrow 0 \quad (\mbox{ as
$\eta\rightarrow\pm\eta_0$}) \; ,                 \label{eq:boundary}
\end{equation}
where $\Ba(\delta V) \equiv \omega_*^2\delta V - \eta\partial\delta
V/\partial\eta$.

We shall also make the physically plausible assumption that the buoyancy
frequency $N_z$ is finite [vanishing on the midplane according to equation
(\ref{eq:buoy})], and remains so as $p \rightarrow 0$ at the boundary. It
then follows that $\Ab$, and therefore also $\Aa$ and $\Ac$, are finite.
From hydrostatic equilibrium, we then obtain limiting expressions for the
following unperturbed functions:
\begin{equation}
p \sim f(r)(\eta_0-\eta)^{\frac{\Gamma}{\Gamma-1}} \; , \quad
g = \frac{(p/p_c)}{(\rho/\rho_c)} \sim - \left(\frac{\Gamma-
1}{\Gamma}\right) \eta_0(\eta-\eta_0)
\label{eq:pgbound}
\end{equation}
as $\eta\rightarrow\pm\eta_0$.

With this behavior near the boundary, the vertical separated equation
(\ref{eq:sep2}) assumes the form $\partial^2 V_\eta / \partial\eta^2 +
\Ba(V_\eta)/(\Gamma g) \rightarrow 0$. However, the boundary condition
(\ref{eq:boundary}) is satisfied if $\rho\Ba(V_\eta)$ vanishes, while the
stronger condition that $\Delta p/p$ be finite requires that
$\Ba(V_\eta)/g$ remains finite at the boundary. These conditions thus
translate into the basic requirement that $\partial^2V_\eta/\partial\eta^2$
be finite at the boundary. Since the boundary is a singular point of the
first kind for equation (\ref{eq:sep2}) (see also section 3.3), it can be
shown that one of the two independent solutions of this equation is regular
at $\eta = \pm\eta_0$, with equivalently $V_\eta$ being finite there.

It is then seen from equations (\ref{eq:sep2}) and (\ref{eq:aux1}) that in
order that these regular solutions remain finite through the disk, we must
require that $N_z^2$ remain less than $\omega^2$ to avoid the `buoyancy
resonance' at $\omega^2=N_z^2$ which occurs in equations (\ref{eq:master2}) and (\ref{eq:aux1}). 

\pagebreak[3]
\section{G-MODE ANALYSIS: EIGENFUNCTIONS}

We are now in a position to attempt to obtain analytical estimates of
properties of the relativistically trapped modes of oscillation of the
disk. We will concentrate on the analog of the gravity (g)--modes, since
they appear to be least sensitive to the poorly known physical conditions at the inner (and outer) radial boundary of the disk. In addition, they typically occupy the largest area of the disk. The pressure (p)--modes and ($m=1$) corrugation modes usually occupy a narrow region bounded by the inner edge of the disk (NW2,3; \markcite{P} Perez 1993; \markcite{K0} Kato 1990; \markcite{I} Ipser 1994).

\begin{figure}
\epsscale{1.0}
\plotone{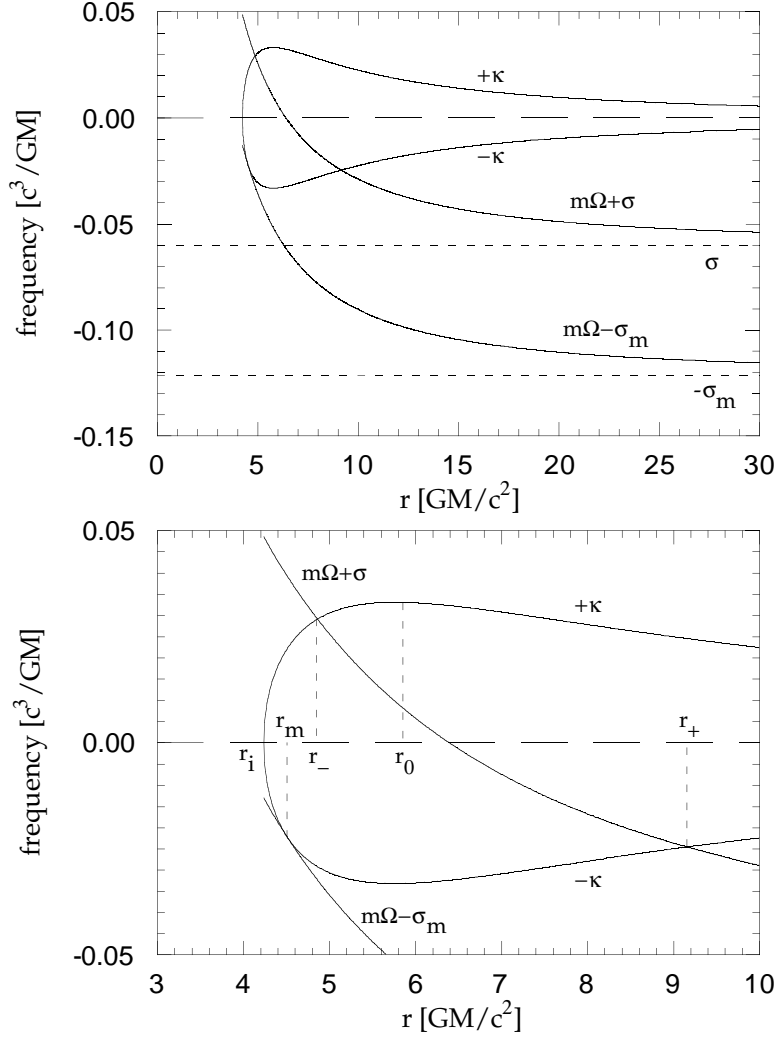}
\caption{The functions which determine the radii $r_-(\sigma)$ and $r_+(\sigma)$ between which the eigenfunctions are concentrated. Also indicated is the maximum value $\sigma_m$ of the eigenfrequency $|\sigma|$, and the inner radius $r_i$ of the disk. For this plot, we choose $a=0.5$ and $m=1$ (with the case $m=0$ corresponding to use of the dashed horizontal lines).}  
\label{function}
\end{figure}

\begin{figure}
\epsscale{1.0}
\plotone{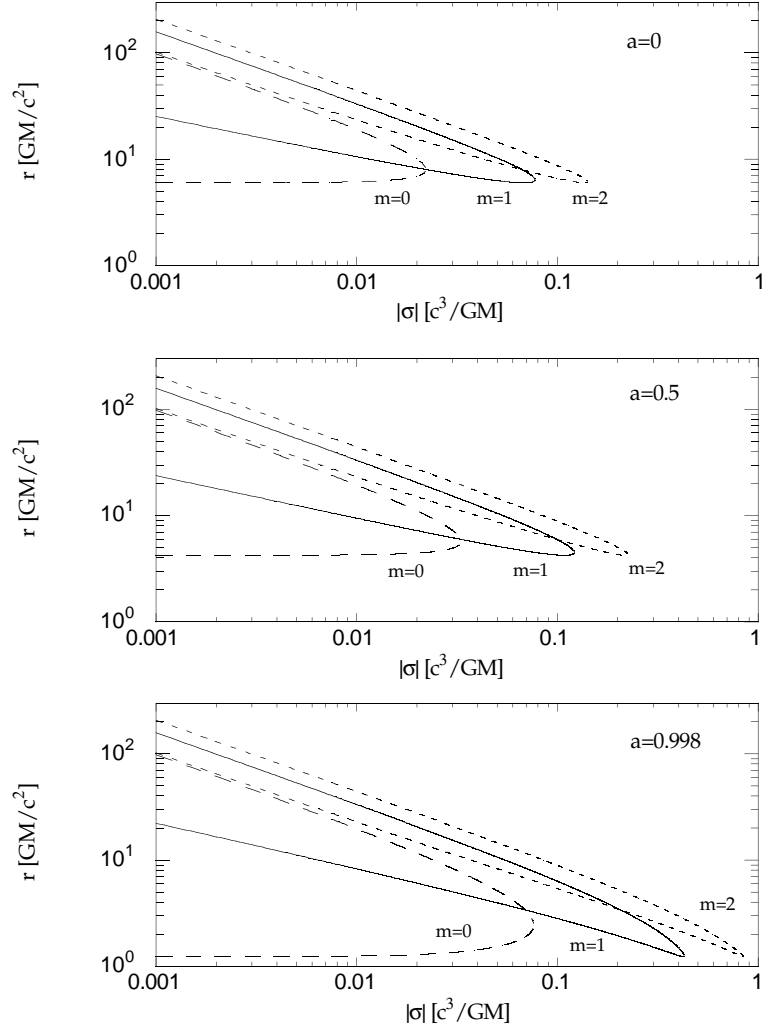}
\caption{The radii between which a g--mode eigenfunction is concentrated is plotted versus its eigenfrequency.}
\label{radii}
\end{figure}

The g--modes are found to exist for values of the separation function
$\Psi$ which satisfy
\begin{equation}
\Psi > \omega_*^2 \quad \quad [\; \omega_* \equiv \omega/\Omega_\perp \; ] \; ,                                                  \label{eq:Psilim} 
\end{equation}
corresponding to $\kappa^2 - \omega^2 > 0$ from equation (\ref{eq:sep1}).
Relativistic gravity forces the radial epicyclic frequency $\kappa$ to
reach a maximum $\kappa_0$ at a radius $r_0$ which is larger than that of the inner edge of the disk $r=r_i$ (very near the innermost stable circular orbit). Therefore we can require that these modes become effectively evanescent at $r<r_-(\sigma)$ and $r>r_+(\sigma)$, where $r_-$ and $r_+ \; (\ge r_-)$ are the places where $\kappa^2(r)=\omega^2(r)$. Recall that the eigenfrequencies $\sigma$ (measured far from the disk) are related to the corotating frequencies by $\sigma = \omega(r) - m\Omega(r)$, and that $\Omega(r) > \kappa(r) > 0$. 
In Figure~\ref{function} we plot these functions that determine the radii $r_\pm(\sigma)$ between which the eigenfunctions are concentrated.
In Figure~\ref{radii} we show the dependence of these boundaries of the trapping region on the eigenfrequency.

We see that for $m=0$ these g--modes exist if $\omega^2 = \sigma^2 < \kappa_0^2$. For $m>0$ they exist if $0 > \sigma > -\max\{ m\Omega(r) + \kappa(r) \} \equiv -(m\Omega_m+\kappa_m)$, with the maximum occuring at $r=r_m$. Since only the square of $\omega$ appears in our fundamental equation (\ref{eq:master2}) for the scalar potential $\delta V$, the case of negative $m$ is obtained by the replacements $m \rightarrow -m$, $\sigma \rightarrow -\sigma$, as is also seen from Figure~\ref{function}. Therefore since only $|\sigma|$ is observable we can assume without loss of generality that $m\ge 0$ and $\sigma<0$ throughout this paper. Thus $0 < -\sigma < \sigma_m \equiv \kappa_m + m\Omega_m$.

The eigenfrequencies are also bounded by the condition
\begin{equation}
\max\{ N_z^2 \} < \omega^2  \; .                        \label{eq:bound}
\end{equation}
This inequality follows from the buoyancy resonance avoidance
discussed at the end of the previous section. The maximum refers to the
behavior of $N_z^2$ as a function of $z$ (for each value of $r$) and is typically expected to be reached at the boundary of the disk.

\pagebreak[3]
\subsection{Radial WKB Solutions}

Within our radial WKB approximation, the radial separated equation
(\ref{eq:sep1}) can be written in the form of the Schr\"{o}dinger equation
\begin{equation}
d^2 W/dr^2 + \Pi (\kappa^2 - \omega^2) W = 0 \; , \label{eq:Schrodinger}
\end{equation}
where $W \equiv (\kappa^2 - \omega^2)^{-1}dV_r/dr$ and $\Pi \equiv
\alpha^2[(\Psi/\omega_*^2) - 1]$. From equations (\ref{eq:xirz}), we see
that the displacement components $\xi^r \propto W$ and $\xi^z \propto V_r =
-\Pi^{-1}dW/dr$.
In the vicinity of $r=r_\mu (\mu=-,+)$, we have $\Pi(\kappa^2-\omega^2)
\cong \Pi_\mu{\cal D}_\mu(r-r_\mu)$ if $r_\mu$ is not close
to $r_m$. (We will treat the case of $r_\mu$ close to $r_m$ later.) Here the various quantities
$Q_\mu \equiv Q(r_\mu)$ and ${\cal D} \equiv d(\kappa^2-\omega^2)/dr$.
[Recall that $\omega = \omega(r)$ is a constant only if $m=0$.]

Near $r_+$, we introduce the coordinate $x_+ = q_+(r-r_+)$, $q_+ \equiv
(\Pi|{\cal D}|)_+^{1/3}$. This transforms equation
(\ref{eq:Schrodinger}) in the vicinity of $r_+$ into the Airy equation
$d^2W/dx_+^2 - x_+W = 0$.
Its solution which exponentially decays for $r>r_+$ is
\begin{equation}
W \propto Ai(x_+) = \pi^{-1}(x_+/3)^{1/2}K_{1/3}(2x_+^{3/2}/3) \; ,
\label{eq:W+}
\end{equation}
where $K_\nu(X)$ [and $I_\nu(X)$] are the modified Bessel functions of
imaginary argument.
For $q_+(r_+-r) \gg 1$, this has the asymptotic form
\begin{equation}
W \propto (r_+ - r)^{-1/4}\cos[(2/3)q_+^{3/2}(r_+ - r)^{3/2} - \pi/4] \; .
\label{eq:W+a}
\end{equation}

Near $r_-$, the governing equation is the same with $x_+$ replaced by $x_-
= q_-(r_- - r)$, $q_- \equiv (\Pi|{\cal D}|)_-^{1/3}$. However,
because the inner edge of the disk (at $r_i$) is not far from $r_-$, we
must now include the other independent Airy function
\[
Bi(x_-) = (x_-/3)^{1/2}[I_{-1/3}(2x_-^{3/2}/3)+I_{1/3}(2x_-^{3/2}/3)] \; ,
\]
so that in the vicinity of $r_-$
\begin{equation}
W \propto Ai(x_-) + \theta Bi(x_-) \; .                  \label{eq:W-}
\end{equation}
The mixing coefficient $\theta$ of the part that grows exponentially for
$r$ well inside $r_-$ is determined from the boundary condition at $r_i$.

The asymptotic forms of this solution are
\begin{equation}
W \propto (r-r_-)^{-1/4}\cos[(2/3)q_-^{3/2}(r-r_-)^{3/2} + (\pi/4) -
\arctan(1/2\theta)]                                    \label{eq:W-a+}
\end{equation}
for $q_-(r-r_-) \gg 1$, and
\begin{equation}
W \propto (r_- - r)^{-1/4}\{ \exp[-(2/3)q_-^{3/2}(r_- - r)^{3/2}] +
2\theta\exp[+(2/3)q_-^{3/2}(r_- - r)^{3/2}] \}           \label{eq:W-a-}
\end{equation}
for $q_-(r_- - r) \gg 1$.
Away from $r_-$ and $r_+$, we may employ the standard WKB approximation,
which takes the form
\begin{equation}
W \propto [\Pi(\kappa^2-\omega^2)]^{-1/4}
\cos\left(\int_{r_-}^r[\Pi(\kappa^2-\omega^2)]^{1/2}dr^\prime
+ \Phi_0\right)                                     \label{eq:WKB12}
\end{equation}
well within the interval \{$r_-,r_+$\}, and
\begin{equation}
W \propto [\Pi(\omega^2-\kappa^2)]^{-1/4}
\cosh\left(\int_{r_i}^r[\Pi(\omega^2-\kappa^2)]^{1/2}dr^\prime
+ \Phi_3\right)                                     \label{eq:WKBi1}
\end{equation}
well inside $r_-$. The phases $\Phi_k$ should not be confused with the
function $\Phi$ given by equation (\ref{eq:Phi}). We also define the phases
\[
\Phi_1 \equiv \int_{r_i}^{r_-}[\Pi(\omega^2-\kappa^2)]^{1/2}dr \; , \quad
\Phi_2 \equiv \int_{r_-}^{r_+}[\Pi(\kappa^2-\omega^2)]^{1/2}dr \; .
\]
We now match these WKB solutions to the appropriate asymptotic forms of our
local solutions.

The first overlap region exists at values of $r$ slightly less than $r_-$,
but such that $(r_- - r) \gg 1/q_-$. We first note that the WKB expression
(\ref{eq:WKBi1}) assumes the form
\[
W \propto (r_--r)^{-1/4}\cosh[(2/3)q_-^{3/2}(r_- - r)^{3/2} + \Phi_1 +
\Phi_3]
\]
as $r \rightarrow r_-$ from below. Comparing this with equation
(\ref{eq:W-a-}), we see that these solutions match if
\begin{equation}
\Phi_1 + \Phi_3 = \half\ln(2\theta) \; .                  \label{eq:theta}
\end{equation}
The second overlap region exists at values of $r$ slightly greater than
$r_-$, but such that $(r-r_-) \gg 1/q_-$. There the WKB expression
(\ref{eq:WKB12}) assumes the form
\[
W \propto (r-r_-)^{-1/4}\cos[(2/3)q_-^{3/2}(r-r_-)^{3/2} + \Phi_0]
\]
as $r \rightarrow r_-$ from above. Comparing this with formula
(\ref{eq:W-a+}), we see that these solutions match if
\begin{equation}
\Phi_0 = (\pi/4) - \arctan(1/2\theta) + n_1\pi \; ,   \label{eq:Phi0}
\end{equation}
where $n_1$ is an integer.
The final overlap region exists at values of $r$ slightly less than $r_+$,
but such that $(r_+ - r) \gg 1/q_+$. There the WKB expression
(\ref{eq:WKB12}) assumes the form
\[
W \propto (r_+ - r)^{-1/4}\cos[(2/3)q_+^{3/2}(r_+ - r)^{3/2} + \Phi_0 + \Phi_2]
\]
as $r \rightarrow r_+$ from below. Comparing this with formula
(\ref{eq:W+a}), we see that these solutions match if
\begin{equation}
\Phi_0 + \Phi_2 = - (\pi/4) + n_2\pi \; ,   \label{eq:Phi02}
\end{equation}
where $n_2$ is another integer.

Combining equations (\ref{eq:theta}), (\ref{eq:Phi0}), and (\ref{eq:Phi02})
then gives us our first fundamental WKB relation between the eigenfrequency
$\sigma = \omega(r) - m\Omega(r)$ and the separation function $\Psi(r)$:
$\Phi_2 - \arctan\{ \exp[-2(\Phi_1 + \Phi_3)]\} = (n + 1/2)\pi$,
where $n$ is also an integer. Although $\Phi_1$ and $\Phi_2$ are equal to
positive definite integrals of known functions of the unperturbed disk,
$\Phi_3$ is related to the still unspecified mixing coefficient $\theta$ by
relation (\ref{eq:theta}).

Note that on the midplane $z=0$, $\Delta p = \delta p + \xi^r\partial
p/\partial r$. Assuming that the first term dominates, requiring that
$\Delta p/p$ remain finite as $r \rightarrow r_i$ means that $dW/dr = -\Pi
V_r \rightarrow 0$. Applying this boundary condition to equation
(\ref{eq:WKBi1}) then requires that $\Phi_3 = 0$, equivalent to $\theta =
\half\exp(2\Phi_1)$, and therefore
\begin{equation}
\Phi_2 - \arctan[\exp(-2\Phi_1)] = (n + 1/2)\pi \; .
\label{eq:Phi2}
\end{equation}
This is the first fundamental (implicit) WKB eigenfrequency relation.

\pagebreak[3]
\subsection{Local Approximation}

We will see that the lowest g--modes (those which have the smallest number of radial nodes in their eigenfunction, corresponding to the highest frequencies) are concentrated in the neighborhood of $r=r_m$. The largest value of $|\sigma|$ occurs where $r_-(\sigma)$ and $r_+(\sigma)$ merge at $r=r_m$, which is where $\omega=-\kappa$ and $d(\kappa^2-\omega^2)/dr = 2\kappa_m(d\kappa/dr+md\Omega/dr) = 0$, as shown in Figure~\ref{function}. Recall that $m \ge 0$, $\sigma < 0$, and quantities evaluated at $r=r_m$ (and, if necessary, $\sigma=-\sigma_m$) have a subscript $m$, with the maximum allowed value of $|\sigma|$ denoted by $\sigma_m = \kappa_m + m\Omega_m$. 
 
Thus it is reasonable to look for an approximate analytic solution $W(r)$ to the reformulated radial equation (\ref{eq:Schrodinger}) via a local approximation of the function $\kappa^2(r) - \omega^2(r)$. Therefore we 
employ the expansion (truncated at first order in $|\sigma|-\sigma_m$ and second order in $r-r_m$) 
\begin{equation}
\kappa^2-\omega^2 \cong 2\kappa_m(\sigma+\sigma_m) - {\cal D}_2 (r-r_m)^2  \; ,                                            \label{eq:local1}
\end{equation}
where ${\cal D}_2 = -(1/2)[d^2(\kappa^2-\omega^2)/dr^2]_m$. We 
ignore the less critical radial dependence of $\Pi = \alpha^2(\Psi/\omega_*^2 - 1)$ in equation (\ref{eq:Schrodinger}), letting $\Pi(r) = \Pi_m$.

If we also employ a dimensionless radial distance $x$ and dimensionless
eigenvalue $\lambda$ defined by $x^4 = \Pi_m{\cal D}_2 (r-r_m)^4$ and $\lambda = 2(\Pi_m/{\cal D}_2)^{1/2}\kappa_m (\sigma+\sigma_m)$, equation (\ref{eq:Schrodinger}) assumes the form of the quantum harmonic oscillator equation $d^2W/dx^2 + (\lambda - x^2)W = 0$. Its eigenfunctions which decay to zero far from $r_m$ and the corresponding eigenvalues are given by the well-known expressions
\begin{equation}
W_n \propto H_n(x)\exp(-\half x^2) \quad , \quad
\lambda = 2n+1 \quad (n=0,1,2,\ldots) \; ;           \label{eq:eigen}
\end{equation}
where $H_n$ are the Hermite polynomials. They may be constructed from the
expression $H_n = (-1)^n\exp(x^2)[d^{\, n}\exp(-x^2)/dx^n]$. We shall later
show that these eigenvalues $\lambda_n$ are the same as those obtained in
this local limit from the general expressions of the previous section, so
we have used the same index $n$.

Thus we obtain the first-order expression 
\begin{equation}
\sigma_m - |\sigma_n| = ({\cal D}_2/\Pi_m)^{1/2}\kappa_m^{-1}(n+1/2)
\label{eq:split} \end{equation}
for the splitting of the eigenfrequencies $\sigma_n$. Note, however, that we have not yet included the splitting produced by the vertical eigenfunctions.
To obtain a first rough estimate of the magnitude of the frequency splitting, define an effective radius of curvature $R_m$ by ${\cal D}_2 \equiv  \kappa_m^2/R_m^2$, and use $c_s \sim h\Omega \sim h\Omega_\perp$ in the expression (\ref{eq:aux1}) for $\alpha^2$ in $\Pi$. We then find that
\begin{equation}
\sigma_m - |\sigma_n| \sim (n+1/2)\Omega_m(h/R)_m[\Psi\Omega_\perp^2/\kappa^2 - 1]_m^{-1/2} \; .  \label{eq:approx}
\end{equation}
Thus the lowest modes of thin disks indeed have eigenfrequencies very close to $\sigma_m$. On the plots in Figure~\ref{maxfreq} we have exhibited the dependence of this key frequency on the black hole angular momentum parameter, for $m=0,1,2,3$. (As with all orbital frequencies, $\sigma_m$ is inversely proportional to the mass of the black hole.)

\pagebreak[3]
\subsection{Vertical WKB Solutions}

We next attempt to obtain, in a way similar to that employed in Section 3.1
for the radial separated equation, solutions to the vertical separated
equation. It has the form
\begin{equation}
\partial^2 V_\eta/\partial\eta^2 + a_1(\eta,r)\partial V_\eta/\partial\eta
+ a_2(\eta,r)V_\eta = 0 \; ,                              \label{eq:Veta}
\end{equation}
with the expressions for the coefficients given in equation
(\ref{eq:sep2}). Here we shall also employ a WKB approximation, which in
this case will be valid if
\[
|\partial\ln V_\eta/\partial\eta| \gg |\partial\ln a_2/\partial\eta|  \sim
|\partial\ln a_1/\partial\eta| \sim 1/(\eta_0 - \eta) \; .
\]
This condition is not as easily satisfied as in the radial case, since the
properties of the unperturbed disk (in $a_1$ and $a_2$) vary much more
rapidly in the vertical direction. However, we may take the separation
function $\Psi$ (in $a_2$) to formally be the required large WKB parameter.
Then the standard WKB analysis gives (away from the boundaries
$\eta=\pm\eta_0$) the approximate solution
\begin{equation}
V_\eta \propto a_2^{-1/4}\exp\left[-\half\int_0^\eta a_1(\eta^\prime)
d\eta^\prime\right]\cos\left[\int_0^\eta \sqrt{a_2(\eta^\prime)}
d\eta^\prime - {\cal I}\pi/2 \right] \; ,
\label{eq:vertWKB}
\end{equation}
where ${\cal I}=0$ (or 1) for $V_\eta$ an even (or odd) function of $\eta$.

In our previous analysis of the behavior of the equilibrium properties of
the disk near its boundary, we isolated the dominant terms, which produce
the following asymptotic behavior of the coefficients in equation
(\ref{eq:Veta}):
\[
a_1 \sim -1/[(\Gamma-1)(\eta_0-\eta)] \; , \quad
a_2 \sim \omega_*^2/[\eta_0(\Gamma-1)(\eta_0-\eta)]
\]
as $\eta\rightarrow\pm\eta_0$. Integrating these functions then gives us
the form of the WKB solution (\ref{eq:vertWKB}) near the boundary:
\begin{equation}
V_\eta \propto (\eta_0-\eta)^{-\frac{(3-\Gamma)}{4(\Gamma-1)}}
\cos\left[q(\eta_0-\eta)^{1/2} - \Theta_0 + {\cal I}\pi/2\right] \; ,
\label{eq:bWKB}
\end{equation}
where $q \equiv 2\omega_*[\eta_0(\Gamma-1)]^{-1/2}$ and $\Theta_0 \equiv
\int_0^{\eta_0} \sqrt{a_2(\eta)} d\eta$.

Next, we obtain the behavior of the exact solution of equation
(\ref{eq:Veta}) near the boundary ($\eta\rightarrow\eta_0$), using the
above limiting expressions for $a_1(\eta)$ and $a_2(\eta)$. The nonsingular
solution to the limiting equation is then found to be
\begin{equation}
V_\eta \propto (\eta_0-\eta)^{-\frac{(2-\Gamma)}{2(\Gamma-
1)}}J_{\left(\frac{2-\Gamma}{\Gamma-1}\right)}[q(\eta_0-\eta)^{1/2}] \; ,
\label{eq:Ve}
\end{equation}
involving a Bessel function of the first kind. Now if $(\eta_0-\eta) \gg
1/q^2$, it assumes the same form as the limiting WKB solution
(\ref{eq:bWKB}), but with a specific phase shift $\pi(3-\Gamma)/[4(\Gamma-
1)]$. Matching the two expressions gives the second fundamental (implicit)
WKB eigenfrequency relation 
\begin{equation}
\Theta_0 \equiv \int_0^{\eta_0} \sqrt{a_2(\eta)} d\eta = \pi\left[j +
\frac{1}{4}\left(\frac{3-\Gamma}{\Gamma-1}\right)
+ \frac{{\cal I}}{2}\right] \; ,
\label{eq:Va}
\end{equation}
where $j$ is an integer. We note that with $\Theta_0$ positive, $j > -(3-
\Gamma)/[4(\Gamma-1)]$ for the even eigenfunctions (${\cal I}=0$) and
$j > -(1+\Gamma)/[4(\Gamma-1)]$ for the odd eigenfunctions (${\cal I}=1$).

As usual, all the results in this section hold for any axial mode number
$m$, which appears in the representation of the corotating frequency
$\omega(r) = \sigma + m\Omega(r)$.

\pagebreak[3]
\section{EIGENVALUES AND EIGENFREQUENCIES}

In our WKB approximation, the vertical eigenvalues $\Psi_j(r,\omega^2)$ of
the separation function $\Psi$ and the corotating eigenfrequencies $\omega = \omega_{nj}(r) = \sigma_{nj} + m\Omega(r)$ are obtained as roots of the radial and vertical
transcendental equations (\ref{eq:Phi2}) and (\ref{eq:Va}), respectively.
We reproduce them in the following forms:
\begin{displaymath}
\int_{r_-}^{r_+}\alpha[(\Psi/\omega_*^2-1)(\kappa^2-\omega^2)]^{1/2}dr
- \arctan\;\exp\left[-2\int_{r_i}^{r_-}\alpha[(\Psi/\omega_*^2-
1)(\omega^2-\kappa^2)]^{1/2}dr\right]
\end{displaymath}
\begin{equation} = (n+1/2)\pi \; , \label{eq:Reigen}
\end{equation}
\begin{equation}
\int_0^{\eta_0}\left[\frac{\Psi-\omega_*^2}{\Gamma}-
\frac{\omega_*^2\partial p/\partial\eta}{\Gamma\eta p}+Q\right]^{1/2} d\eta
= (j + \delta_{\Gamma})\pi \; ,                          \label{eq:Veigen}
\end{equation}
where we require $n \geq 0$ and $j+\delta_{\Gamma} > 0$, with
\begin{equation}
\delta_{\Gamma} = \frac{3-\Gamma}{4(\Gamma-1)} \quad \mbox{ (even modes) },
\quad \frac{1+\Gamma}{4(\Gamma-1)} \quad \mbox{ (odd modes) } \; .
\end{equation}
We have also introduced the quantity
\begin{equation}
Q(\eta, r, \Psi, \omega_*^2) =  \left[\frac{\eta}{\Gamma}
\left(\frac{\Psi}{\omega_*^2}-1\right) - \Xi\right]\frac{\partial\Aa
}{\partial\eta} + \frac{\partial^2\Aa}{\partial\eta^2} \; .
\end{equation}
Note that $\alpha^2(r) = \beta^2g_{rr}\rho_c/(\Gamma p_c)$.
Recall the restriction (\ref{eq:Psilim}) on the separation function
$\Psi_j$ and the lower limit (\ref{eq:bound}) and the upper limits on the eigenfrequencies given in the previous section.

Equation (\ref{eq:Reigen}) implies that
\begin{equation}
(n+1/2)\pi \le \int_{r_-}^{r_+}\alpha[(\Psi_j/\omega_*^2-1)(\kappa^2-
\omega^2)]^{1/2}dr \le (n+1)\pi   \; .    \label{eq:limit1}
\end{equation}
Since for any given value of $j$, $\Psi_j(r,\omega)$ is bounded everywhere
in the (finite) interval $r_- \le r \le r_+$, as are all the other
quantities in the above integrand, the first inequality shows that $n$ has
a finite maximum value. Thus for any given value of $j$, there exist only a
finite number of eigenfrequencies $\sigma_{nj}$.
On the other hand, equation (\ref{eq:Veigen}) then shows that
$\Psi_j\rightarrow \infty$ for $j\rightarrow\infty$. Therefore, for any
given value of $n \ge 0$, the second inequality in equation
(\ref{eq:limit1}) implies that $|\sigma_{nj}|\rightarrow\kappa_m+m\Omega_m \equiv \sigma_m$ (for any $m \ge 0$) from below (with $r_\pm \rightarrow r_m \pm 0$) as $j\rightarrow\infty$.

Using the fact that the left hand side of equation (\ref{eq:Reigen}) is a
continuous function of $\omega$ it follows that for a given value of $n$
this equation has at least one root $\sigma_{nj}$ for any sufficiently
large value of $j$. We can therefore summarize the above results as follows.

The spectrum of eigenfrequencies $\sigma_{nj}$ in the range $-m\Omega(r_i) > \sigma > -[m\Omega(r_m)+\kappa(r_m)] = -\sigma_m$ is discrete, with the only condensation point $\sigma_m$. For any given integer $j$ satisfying $j + \delta_\Gamma > 0$, the number of eigenfrequencies is finite, with $n = n_{min}^{(j)}, n_{min}^{(j)}+1, \ldots , n_{max}^{(j)}$. For any given integer $n \ge 0$, there is a sequence of eigenfrequencies which satisfies
\begin{equation}
\lim_{j\rightarrow\infty} \sigma_{nj} = -(\kappa_m +m\Omega_m) \equiv -\sigma_m  \; .                                      \label{eq:eigenlim} 
\end{equation}
For $m=0$, either sign is allowed, but for convenience we have chosen the negative.
Typically, there will be monotonicity in $n$ and $j$; that is,
$ |\sigma_{n+1,j}| < |\sigma_{n,j}| $ and $ |\sigma_{n,j+1}| > |\sigma_{n,j}| $. 

\pagebreak[3]
\subsection{Local Approximation}

For eigenfrequencies close to $\sigma_m$, with $\kappa^2(r)-\omega^2(r)$ approximated by equation (\ref{eq:local1}), the first term on the left hand side of equation (\ref{eq:Reigen}) becomes small as $r_-, r_+ \rightarrow r_m$, but the second one is exponentially small and thus may be neglected. After some algebra, this equation then provides the approximate expression
\begin{equation}
\sigma_m - |\sigma_{nj}| \cong \frac{\sqrt{{\cal D}_2}}{\kappa_m\alpha(r_m)} \left[\frac{\Psi_j(r_m,\kappa_m)\Omega_\perp^2(r_m)}{\kappa_m^2} - 1 \right]^{-1/2} \left(n+\frac{1}{2}\right) \; .                    \label{eq:local2}
\end{equation}
This expression is consistent with the previous quantization
(\ref{eq:eigen}) of the eigenvalue $\lambda_n$ obtained from a more
specific analysis of the local approximation, as can be seen from the
definition of $\lambda$ in that section.

Using this expression and the fact that $\Psi_j \propto j^2$ for large $j$
from equation (\ref{eq:Veigen}), we see that $\sigma_m - |\sigma_{nj}|
\propto (n+1/2)/j$ for $j\rightarrow\infty$.

\pagebreak[3]
\subsection{Small Eigenfrequencies (for $m=0$)}

Let us now investigate the low frequency regime of the radial ($m=0$) modes, in which $r_- \rightarrow r_i$ and $r_+ \rightarrow \infty$, which is in a sense opposite to that of the local approximation. As $\omega = \sigma \rightarrow 0$, the first term in equation
(\ref{eq:Reigen}) becomes
\begin{equation}
\int_{r_-}^{r_+}\alpha(r)[(\Psi/\omega_*^2-1)(\kappa^2-\omega^2)]^{1/2}dr
\cong |\omega|^{-1}\int_{r_i}^{\infty}
\alpha\Psi^{1/2}(r,\omega)\Omega_{\perp}\kappa\, dr \equiv
\Upsilon(\omega)/|\omega| \; ,  \label{eq:upsilon}
\end{equation}
while the second term becomes $-\pi/4$, so this equation then gives
\begin{equation}
|\sigma_{nj}| = \frac{\Upsilon_j(\sigma_{nj})}{(n+3/4)\pi} \ll \kappa_0 \quad (m=0) \; .                 \label{eq:lowfreq}
\end{equation}
However, this limiting behavior can only exist if the buoyancy frequency
$N_z$ is also small, since it provides a lower limit to $\omega$ as
indicated in equation (\ref{eq:bound}). In addition, it is seen that
$\Upsilon_j \sim \Psi_j^{1/2}(r/h)_0\Omega_0$, so this limiting
behavior of $\sigma_{nj}$ can be reached only for very large values of $n$.

\pagebreak[3]
\subsection{Neutrally Buoyant Disks}

The vertical eigenfunction equation (\ref{eq:sep2}) and eigenvalue equation
(\ref{eq:Veigen}) simplify greatly if the buoyancy frequency $N_z$
($\propto \Ab$) vanishes. Equation (\ref{eq:A}) shows that this will occur
if the specific entropy $s$ is only a function of density, giving the
barotropic equation of state $p = p(\rho, s) = p(\rho)$. In practice, this
can usually only be achieved by the entropy being constant (in the $z$
direction), such as within convective zones of stars.

The relation between pressure and density in the equilibrium (unperturbed)
disk is determined by the equation of state, the radiative and turbulent
advective transport of energy, and the viscous energy generation rate. The
last two poorly known functions are critical in establishing the vertical
gradient of the specific entropy $s$. However, as shown by Nowak \& Wagoner
\markcite{NW5} (1995), with the maximum optical depths of accretion disks much less than within stars, advection can never dominate the transfer of energy.
Therefore, the vertical gradient of the entropy cannot vanish. We also do
not expect that $s = s(\rho)$, so the buoyancy frequency should be nonzero, although it could be relatively small.

Nevertheless, we shall now assume that the buoyancy frequency vanishes, in
order to obtain relatively simple results that can be used as a guide to
the nature of the eigenvalues $\Psi_j$ and eigenfrequencies $\sigma_{nj}$.
In the next section, we relax this assumption to small buoyancy
frequencies.

When $\Ab=0$, equation (\ref{eq:A}) gives $p(r,z) = F_p(r)\rho^\Gamma$.
From vertical hydrostatic equilibrium [equation (\ref{eq:hydrostatic})] one
then obtains
\begin{equation}
p = p_c(r)\left(1-\frac{\eta^2}{\eta_0^2}\right)^{1/\zeta} \, , \quad
\eta_0^2 = \frac{2}{\zeta} \equiv \frac{2\Gamma}{\Gamma-1} \; .
\label{eq:pressure}
\end{equation}
(We note that the corresponding expression for $\eta_0^2$ in NW3 lacks the
factor of 2.) With $Q=0$, equation (\ref{eq:Veigen}) then gives the
vertical eigenvalue relation
\begin{equation}
\Psi_j = \frac{\pi^2(\Gamma-1)}{2E^2(k_{nj})}(j+\delta_\Gamma)^2 \; ,
\quad\quad  k_{nj}^2 = 1 - \frac{\omega_{nj}^2}{\Omega_\perp^2\Psi_j} \; ,
\label{eq:isen}
\end{equation}
where $E(k)$ is the complete elliptic integral of the second kind.
Employing the local approximation, equations (\ref{eq:local2}) and (\ref{eq:isen}) allow the determination of $\Psi_j$ and $\omega_{nj} = \sigma_{nj} + m\Omega_m$ for large eigenfrequencies.

A graphical solution is presented in Figure~\ref{graphPsi}, obtained in the
following manner. Substituting equation (\ref{eq:local2}) for $\sigma_{nj}$
into the expression for $k_{nj}(\omega_{nj})$ in equation (\ref{eq:isen}) gives the function $k_{nj} = f_1(\Psi_j; n, N_1, N_2)$, where the dimensionless parameters $N_1(a) = \kappa_m/\Omega_\perp \sim 1$ and $N_2(c_s(r_m,0)/c,a) = \sqrt{{\cal D}_2}/(\kappa_m\Omega_\perp\alpha_m) \sim h_m/R_m$. Then equation (\ref{eq:isen}) can be written in the form
\begin{equation}
(j+\delta_\Gamma)^{-2}\Psi_j = f_2(\Psi_j; n, \Gamma, N_1, N_2) \; .
\label{eq:PsiPlot}
\end{equation}
Therefore, in Figure~\ref{graphPsi} we have plotted [for a specific choice
of $a$, $\Gamma$, and $c_s(r_m,0)$] each side of equation
(\ref{eq:PsiPlot}) versus $\Psi_j$, for various values of $j$ (left
side), $n$ (right side), and $m$. From their intersections we obtain the eigenvalues $\Psi_j$. Then from equation (\ref{eq:local2}) we obtain the
eigenfrequencies $\sigma_{nj}$. 

In Tables~\ref{a0}, \ref{ahalf}, and \ref{a1} we present (for three choices of $a$ as well as $m$) the eigenvalues and corresponding eigenfrequencies for radial mode numbers $n\le 1$, even vertical mode numbers $j\le 1$, and odd vertical mode numbers $j\le 0$. For comparison, we also indicate the corresponding values of the maximum eigenfrequency $\sigma_m$. 

The expression (\ref{eq:isen}) provides the following bounds on the
eigenvalue, since $E(0)=\pi/2$ and $E(1)=1$:
\begin{equation}
2(\Gamma-1)(j+\delta_\Gamma)^2 \le \Psi_j \le \case{1}{2}\pi^2(\Gamma-1)
(j+\delta_\Gamma)^2 \; .
\end{equation}
$E(k)$ is also monotonic between 0 and 1, so the graphical solution shows
that for any allowed value of $j$ there is exactly one eigenvalue $\Psi_j$.
For large enough values of $j$, it is close to its upper bound: $\Psi_j
\approx \case{1}{2}\pi^2(\Gamma-1) (j+\delta_\Gamma)^2 + O(1)$.

In the low-frequency limit (for $m=0$), $\sigma_{nj}^2 \ll \kappa_0^2$, equation (\ref{eq:isen}) gives $\Psi_j \cong [(\Gamma-
1)/2]\pi^2(j+\delta_\Gamma)^2$, while equations(\ref{eq:upsilon}) and
(\ref{eq:lowfreq}) give
\begin{equation}
\omega_{nj} \cong \left(\frac{j+\delta_\Gamma}{n+3/4}\right)
\left(\frac{\Gamma-1}{2}\right)^{1/2} \Upsilon^{(o)} \; , \quad
\Upsilon^{(o)} \equiv \int_{r_i}^{\infty} \alpha\Omega_{\perp}\kappa\, dr
\; .
\end{equation}

\pagebreak[3]
\subsection{Nearly Neutrally Buoyant Disks}

We now consider the case of buoyancy frequencies $N_z$ which are finite,
but still much smaller than the characteristic frequency $\Omega$. Equation
(\ref{eq:pressure}) then is modified to the form
\begin{equation}
p = p_c(r)(1-\eta^2/\eta_0^2)^{1/\zeta}[1-f(\eta)] \; ,
\end{equation}
where $f(0)=0$ and $|f(\eta)| \ll 1$. The vertical eigenvalue equation
(\ref{eq:Veigen}) takes a particularly simple form if we choose $f(\eta) =
\case{1}{2}f_0\eta^2$, with $f_0$ a constant:
\begin{equation}
\Psi_j \cong \frac{(\Gamma-1)(j+\delta_\Gamma)^2\pi^2}{2E^2(k_{nj})} \left\{1 + \frac{\omega_*^2f_0}{\Gamma}\left[\left(1-\frac{1}{k_{nj}^2}\right)K(k_{nj}) +
\frac{1}{k_j^2}E(k_{nj})\right]\right\} \; ,
\end{equation}
with $K(k)$ the complete elliptic integral of the first kind. Note that
this result represents the first-order correction to that of the
neutrally-buoyant disk for any slowly-varying (even) function $f(\eta)$.

\pagebreak[3]
\section{CONCLUSIONS}

The most observable g--modes will be those with the lowest values of the
radial index $n$, since the radial compressions and rarefactions of the
disk gas density and temperature will tend to produce the greatest
modulations of the total luminosity . The luminosity modulations of modes
with large values of the vertical index $j$ should be less suppressed.
These low $n$ (high-frequency) modes are also those for which the local
approximation is most valid. Therefore we can use equation (\ref{eq:eigen}) to estimate (taking $x \sim 1$) that the effective radial width $\Delta r = r_+-r_-$ of the
eigenfunction is of order
\begin{equation}
\Delta r \sim \sqrt{hR_m} \propto \sqrt{c_s(r_m,0)} \; .
\end{equation}
Recall that $R_m$ is an effective radius of curvature of the function $\kappa^2-\omega^2$, defined in Section 3.2. We find that for $m=0$, 
$R_0/r_0$ varies from 0.408 for $a=0.0\; (r_0=8.000)$ to 0.660 for $a=0.998\; (r_0=2.449)$. For $m=2$, $R_2/r_2$ varies from 0.0444 for $a=0.0\; (r_2=6.141)$ to 0.00880 for $a=0.998\; (r_2=1.242)$.

We can also use equations (\ref{eq:xirz}) and (\ref{eq:xiphi}) to obtain
the following estimates of the relative magnitudes of the components of the
fluid diplacement vector $\vec{\xi}$, in a local orthonormal basis, in terms of the scalar potentials $\delta V = V_rV_\eta = \delta p/(\rho\beta\omega)$ and $W = (\kappa^2 - \omega^2)^{-1}dV_r/dr$:
\begin{equation}
\frac{\xi^{\hat{z}}}{\xi^{\hat{r}}} \sim
\left(\frac{\Omega}{\omega}\right)^2
\left(\frac{h}{V_\eta}\frac{\partial V_\eta}{\partial z}\right)
\left(\frac{h}{W}\frac{dW}{dr}\right) \; , \quad \quad
\frac{\xi^{\hat{\phi}}}{\xi^{\hat{r}}} \sim \frac{\Omega}{\omega} \; .
\end{equation}
We also note that there are $n$ radial nodes (corresponding to
$\omega_{nj}$) in $W(r)$, $j+1$ vertical nodes (corresponding to $\Psi_j$)
in even $V_\eta(\eta,r)$ for $11/9<\Gamma\le 7/5$ and in odd
$V_\eta(\eta,r)$ for $9/7<\Gamma\le 5/3$, and $j$ vertical nodes in even
$V_\eta(\eta,r)$ for $7/5<\Gamma\le 3$ and in odd $V_\eta(\eta,r)$ for
$5/3<\Gamma\le\infty$. Only vertical nodes in the interval $0 < \eta < \eta_0$ have been counted. Since the pressure perturbation $\delta p \propto V_rV_\eta \propto (dW/dr)V_\eta$, it will instead have $n+1$ radial nodes.

The accretion disk model chosen for Figures~\ref{size} and \ref{graphPsi} and for Tables~\ref{a0}, \ref{ahalf}, and \ref{a1} is radiation-dominated in the region where the mode is trapped, corresponding to $\Gamma=4/3$. There are thus $j+1$ vertical nodes in both even and odd vertical eigenfunctions. From the results in these tables, we therefore see that the lowest even and odd temperature perturbations $\delta T \propto \delta p$ have one radial node. On the other hand, $\delta T$ has no vertical node for the lowest odd mode. The lowest even mode has one vertical node for low values of $a$ and $m$, and no vertical node for the higher values. (Note, however, that the total actual number of vertical nodes will be twice as many for the even modes and one plus twice as many for the odd modes.)

The eigenfrequencies $-\sigma_{nj}$ of the low $n$ modes lie just below the maximum frequency $\sigma_m$. Using equations (\ref{eq:local2}) and (\ref{eq:isen}), the definition $R_m \equiv \kappa_m/\sqrt{{\cal D}_2}$, and the approximation $\Psi_j\Omega_\perp^2/\omega_{nj}^2 \gg 1$ (which is seen to be reasonably valid from Tables~\ref{a0}--\ref{a1}), the frequency separation is found to be approximated by 
\begin{equation}
\frac{\sigma_m - |\sigma_{nj}|}{\kappa_m} \approx \left[\frac{2\Gamma}{g_{rr}(\Gamma-1)}\right]^{1/2} \frac{(n+1/2)}{(j+\delta_\Gamma)}\left[\frac{h(r_m)}{\pi R_m}\right]
 \; .  \label{eq:freq2}
\end{equation}
For the accretion disk model chosen for Figures~\ref{size} and \ref{graphPsi} and Tables~\ref{a0}--\ref{a1}, corresponding to a typical disk luminosity, we see from these tables that the eigenfrequencies $|\sigma_{nj}|$ of the lowest modes are indeed very close to the maximum g--mode frequency $\sigma_m$. This small splitting  corresponds to a very thin disk [ $h(r_m)/r_m = 6.6\times10^{-4}$ for $a=m=0$ ], as indicated in equation (\ref{eq:freq2}).  
 
\begin{figure}
\epsscale{1.0}
\plotone{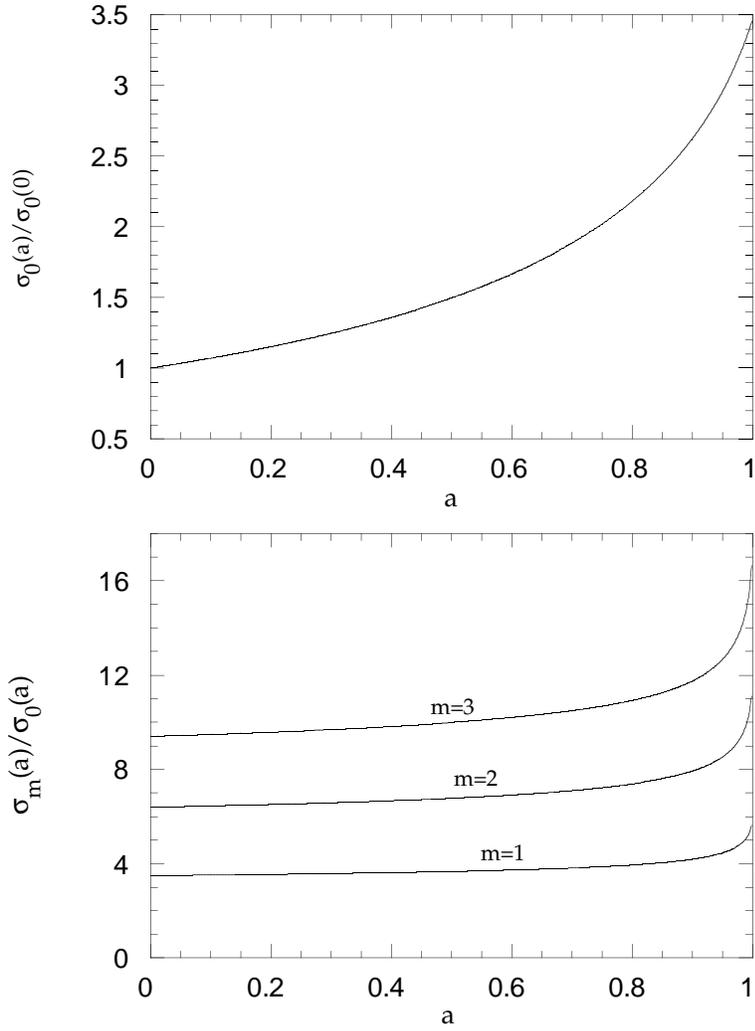}
\caption{(a) The dependence of the maximum radial ($m=0$) eigenfrequency on the black hole angular momentum parameter $a=cJ/GM^2$, relative to its value at $a=0$.
(b) The ratio of the maximum eigenfrequency of some higher $m$ modes to that of the radial mode.}
\label{maxfreq}
\end{figure}

\begin{figure}
\epsscale{1.0}
\plotone{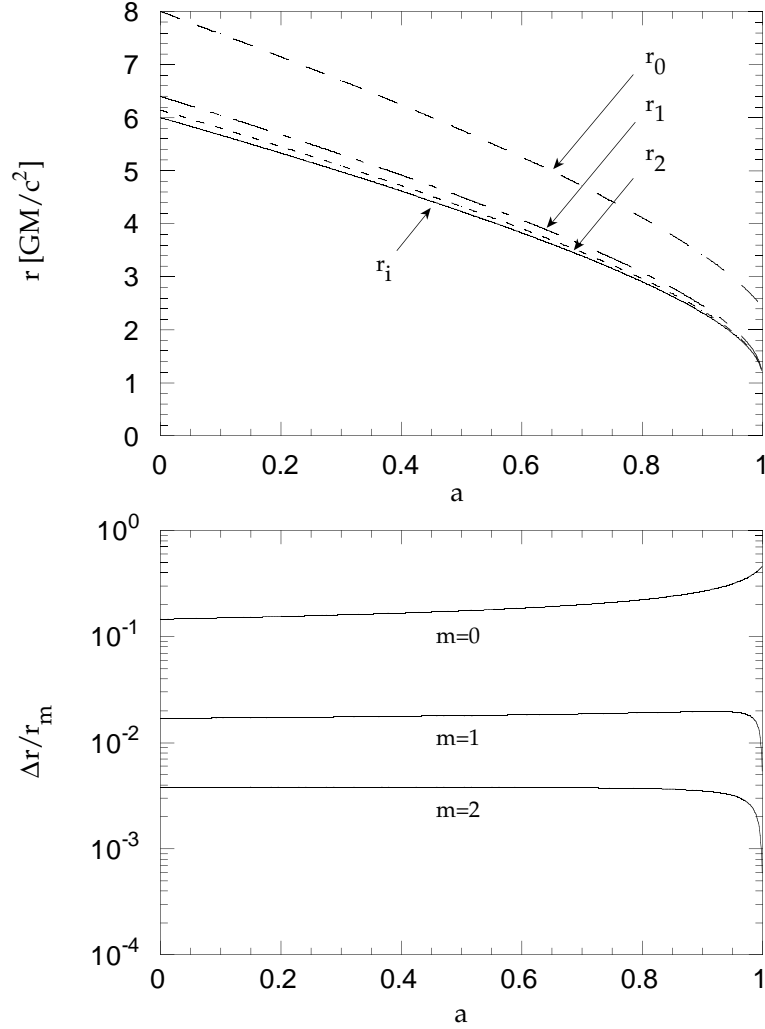}
\caption{(a) The black hole angular momentum dependence of the radius $r_m$ to which $r_-$ and $r_+$ converge as $|\sigma| \rightarrow \sigma_m$. Also shown is the radius $r_i$ of the inner edge of the disk.
(b) The dependence of the fractional effective width of the lowest eigenfunction, $\Delta r/r_m = [r_+(\sigma) - r_-(\sigma)]/r_m$, on the angular momentum of the black hole. The same values of $m$ are chosen as in (a). The accretion disk model is specified by $\Gamma=4/3$, a locally isentropic equation of state, and speed of sound corresponding to a luminosity $L=0.1L_{Edd}$ from a radiation-dominated disk.}
\label{size}
\end{figure}

\begin{figure}
\epsscale{1.0}
\plotone{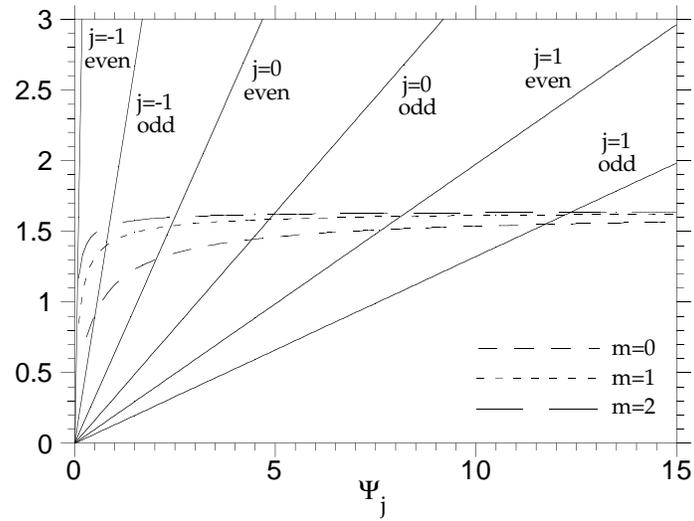}
\caption{Graphical determination of the lowest eigenvalues $\Psi_j$,
assuming the local approximation for a neutrally buoyant disk.
The accretion disk model is the same one used in the previous figure, and we have chosen $a=0.5$. The curves are the right-hand-side of equation (4-12) for $n=0,1,2$ (difference invisible) and the values of $m$ indicated, while the straight lines are the left-hand-side of equation (4-12).}
\label{graphPsi}

\end{figure}

Therefore the low-$n$ g--mode inertial frequencies $f=-\sigma/2\pi$ (now in usual units) of thin disks are essentially a function of only the mass and angular momentum of the black hole and the axial index $m$. For instance, we obtain
\begin{equation}
f(m=0) \cong (c^3/GM)\kappa_0/2\pi = (64\pi\sqrt{2})^{-1}(c^3/GM)F(a) =
714(M_\odot/M)F(a) \mbox{ Hz} \; ,      
\end{equation}
where $F(0)=1$ and $F(a)$ is the upper plot of Figure~\ref{maxfreq}. It is seen that $F(a_{max}) \cong 3.443$, if $a_{max} \cong 0.998$. In order to determine both $M$ and $a$, one needs to observe another eigenfrequency with a different dependence on $a$. 

For instance, in principle one could determine $a$ via the frequency ratio 
\begin{equation}
f(m\neq0)/f(m=0)\cong\sigma_m/\sigma_0 = (\kappa_m+m\Omega_m)/\kappa_0 \; ,
\end{equation} 
which is shown in the lower plot of Figure~\ref{maxfreq} for $m=1,2,3$. Unfortunately, these ratios are close to their $a=0$ values unless $a$ is fairly close to unity. And of course, any $m \neq 0$ mode will not be observable if the disk is viewed face-on, and in general should produce less luminosity modulation (via Doppler boosting, etc.) than the $m=0$ mode.

Finally, we note that for $a \lesssim 0.1$ these results will also apply for accretion disks around slowly rotating neutron stars, since their external metric is Kerr through first order in $a$. However, the requirement that the disk be unaffected means that the neutron star must also be weakly magnetized ($B\lesssim 10^7$ gauss) and compact (soft equation of state). Fully relativitic numerical calculations (\markcite{FI} Friedman \& Ipser 1992; \markcite{CST} Cook, Shapiro, \& Teukolsky 1994) indicate that the maximum value of $a \cong 0.6 - 0.8$ over a wide range of equations of state. Typically, this maximum does not occur at the maximum (uniform) angular velocity.

In future papers, we will extend this relativistic analysis in at least three directions:

(a) We will investigate the characteristics of the other classes of modes,
extending previous analyses of acoustic and corrugation modes. The
dependence of the eigenfrequencies on the angular momentum parameter $a$ is
especially important, as indicated above. For example, the observation of three eigenfrequencies which have significantly different dependences on $a$ would allow one to test the validity of the Kerr metric.  

(b) Numerical solutions for the eigenfunctions and eigenvalues will be
obtained, in a manner similar to that employed in the pseudo-Newtonian
analyses of NW1,2,3. Initially, the radial--WKB separated equations
(\ref{eq:sep1}) and (\ref{eq:sep2}) will be employed.

(c) The assumption of thin disks will be relaxed. This will introduce many
complications; such as noncircular flow, inclusion of more terms in the metric, and the probable need for a fully two-dimensional integration of the Ipser--Lindblom master perturbation equation.

\acknowledgments

This research was supported in part by NASA grant NAG 5-3102 to R.V.W.,
NASA grant NAS8--39225 to Gravity Probe B, and a National Defense Science and Engineering Graduate Fellowship to D.E.L. We thank Lee Lindblom and
Michael Nowak for useful discussions.

\newpage
\begin{table}[t]
\begin{center}
\begin{tabular}{ccrlcc}
$m$ & $n$ & $j$ &  & $\Psi_j$ & $-\sigma_{nj}$ \\
\tableline
0 & 0 & 0 & even & 2.043865 & 0.02196301 \\
0 & 1 & 0 & even & 2.053849 & 0.02169597 \\
0 & 0 & 1 & even & 7.661118 & 0.02203112 \\
0 & 1 & 1 & even & 7.667663 & 0.02189928 \\
0 & 0 & -1 & odd & 0.520825 & 0.02175202 \\
0 & 1 & -1 & odd & 0.538491 & 0.02109409 \\
0 & 0 & 0 & odd &  4.431470 & 0.02200927 \\
0 & 1 & 0 & odd &  4.439250 & 0.02183388 \\ [0.15in]
1 & 0 & 0 & even & 2.396927 & 0.07718115 \\
1 & 1 & 0 & even & 2.397778 & 0.07713537 \\
1 & 0 & 1 & even & 8.118458 & 0.07719172 \\
1 & 1 & 1 & even & 8.119028 & 0.07716707 \\
1 & 0 & -1 & odd & 0.782423 & 0.07716282 \\
1 & 1 & -1 & odd & 0.783638 & 0.07708047 \\
1 & 0 & 0 & odd &  4.843951 & 0.07718805 \\
1 & 1 & 0 & odd &  4.844628 & 0.07715606 \\ [0.15in]
2 & 0 & -1 & even & 0.063096 & 0.14133885 \\
2 & 1 & -1 & even & 0.063578 & 0.14125715 \\
2 & 0 & 0 & even &  2.495455 & 0.14137481 \\
2 & 1 & 0 & even &  2.495589 & 0.14136430 \\
2 & 0 & 1 & even &  8.239465 & 0.14137718 \\
2 & 1 & 1 & even &  8.239554 & 0.14137142 \\
2 & 0 & -1 & odd &  0.861875 & 0.14137104 \\
2 & 1 & -1 & odd &  0.862065 & 0.14135301 \\
2 & 0 & 0 & odd &   4.955285 & 0.14137634 \\
2 & 1 & 0 & odd &   4.955391 & 0.14136890
\end{tabular}
\end{center}
\caption{Eigenvalues and eigenfrequencies of the lowest g--modes, calculated within the local approximation in neutrally buoyant disks. The angular momentum parameter $a=0$, and the accretion disk model is the same as specified in Figure 5. For comparison, the maximum frequencies are $\sigma_0 = 0.0220971$, $\sigma_1 = 0.0772040$, and $\sigma_2 = 0.1413801$. As usual, the frequencies are expressed in units of $c^3/GM$.}
\label{a0}
\end{table}

\newpage
\begin{table}[t]
\begin{center}
\begin{tabular}{ccrlcc}
$m$ & $n$ & $j$ &  & $\Psi_j$ & $-\sigma_{nj}$ \\
\tableline
0 & 0 & 0 & even & 2.034107 & 0.03287054 \\
0 & 1 & 0 & even & 2.046763 & 0.03237219 \\
0 & 0 & 1 & even & 7.647255 & 0.03299819 \\
0 & 1 & 1 & even & 7.655551 & 0.03275271 \\
0 & 0 & -1 & odd & 0.515535 & 0.03246458 \\
0 & 1 & -1 & odd & 0.537941 & 0.03123175 \\
0 & 0 & 0 & odd &  4.419339 & 0.03295735 \\
0 & 1 & 0 & odd &  4.429199 & 0.03263055 \\ [0.15in]
1 & 0 & 0 & even & 2.402442 & 0.12146291 \\
1 & 1 & 0 & even & 2.403417 & 0.12138527 \\
1 & 0 & 1 & even & 8.125275 & 0.12148081 \\
1 & 1 & 1 & even & 8.125928 & 0.12143896 \\
1 & 0 & -1 & odd & 0.786847 & 0.12143202 \\
1 & 1 & -1 & odd & 0.788236 & 0.12129279 \\
1 & 0 & 0 & odd &  4.850206 & 0.12147458 \\
1 & 1 & 0 & odd &  4.850981 & 0.12142028 \\ [0.15in]
2 & 0 & -1 & even & 0.065171 & 0.22445146 \\
2 & 1 & -1 & even & 0.065678 & 0.22432569 \\
2 & 0 & 0 & even &  2.499892 & 0.22450645 \\
2 & 1 & 0 & even &  2.500037 & 0.22448958 \\
2 & 0 & 1 & even &  8.244807 & 0.22451026 \\
2 & 1 & 1 & even &  8.244903 & 0.22450099 \\
2 & 0 & -1 & odd &  0.865544 & 0.22450043 \\
2 & 1 & -1 & odd &  0.865750 & 0.22447153 \\
2 & 0 & 0 & odd &   4.960238 & 0.22450891 \\
2 & 1 & 0 & odd &   4.960353 & 0.22449696
\end{tabular}
\end{center}
\caption{Same as previous table, for angular momentum parameter $a=0.500$. For comparison, the maximum frequencies are $\sigma_0 = 0.0331210$, $\sigma_1 = 0.1215017$, and $\sigma_2 = 0.2245149$.}
\label{ahalf}
\end{table}

\newpage
\begin{table}[t]
\begin{center}
\begin{tabular}{ccrlcc}
$m$ & $n$ & $j$ &  & $\Psi_j$ & $-\sigma_{nj}$ \\
\tableline
0 & 0 & 0 & even & 1.981751 & 0.07467037 \\
0 & 1 & 0 & even & 2.015375 & 0.07189831 \\
0 & 0 & 1 & even & 7.569936 & 0.07539989 \\
0 & 1 & 1 & even & 7.592026 & 0.07404846 \\
0 & 0 & -1 & odd & 0.494041 & 0.07197436 \\
0 & 1 & -1 & odd & 0.551344 & 0.06523649 \\
0 & 0 & 0 & odd &  4.352452 & 0.07517024 \\
0 & 1 & 0 & odd &  4.378675 & 0.07336520 \\ [0.15in]
1 & 0 & -1 & even & 0.071106 & 0.42744608 \\
1 & 1 & -1 & even & 0.073033 & 0.42691367 \\
1 & 0 & 0 & even &  2.511019 & 0.42767939 \\
1 & 1 & 0 & even &  2.511625 & 0.42759925 \\
1 & 0 & 1 & even &  8.258039 & 0.42769742 \\
1 & 1 & 1 & even &  8.258439 & 0.42765333 \\
1 & 0 & -1 & odd &  0.874910 & 0.42765112 \\
1 & 1 & -1 & odd &  0.875766 & 0.42751453 \\
1 & 0 & 0 & odd &   4.972562 & 0.42769104 \\
1 & 1 & 0 & odd &   4.973041 & 0.42763418 \\ [0.15in]
2 & 0 & -1 & even & 0.092114 & 0.84564934 \\
2 & 1 & -1 & even & 0.092263 & 0.84559530 \\
2 & 0 & 0 & even & 2.552778 & 0.84567138 \\
2 & 1 & 0 & even & 2.552826 & 0.84566133 \\ 
2 & 0 & 1 & even & 8.307546 & 0.84567362 \\
2 & 1 & 1 & even & 8.307577 & 0.84566805 \\
2 & 0 & -1 & odd & 0.910007 & 0.84566797 \\
2 & 1 & -1 & odd & 0.910077 & 0.84565112 \\
2 & 0 & 0 & odd &  5.018748 & 0.84567282 \\
2 & 1 & 0 & odd &  5.018786 & 0.84566566 
\end{tabular}
\end{center}
\caption{Same as previous table, for angular momentum parameter $a=0.998$. For comparison, the maximum frequencies are $\sigma_0 = 0.0760771$, $\sigma_1 = 0.4277195$, and $\sigma_2 = 0.8456764$.}
\label{a1}
\end{table}

\newpage

\newpage

\newpage

\newpage

\newpage

\newpage

\end{document}